# Exploring convolutional neural networks with transfer learning for diagnosing Lyme disease from skin lesion images

Sk Imran Hossain[a], Jocelyn de Goër de Herve[b,c], Md Shahriar Hassan[a], Delphine Martineau[d], Evelina Petrosyan[d], Violaine Corbin[d], Jean Beytout[e], Isabelle Lebert[b,c], Jonas Durand[f], Irene Carravieri[g], Annick Brun-Jacob[f], Pascale Frey-Klett[h], Elisabeth Baux[i], Céline Cazorla[j], Carole Eldin[k], Yves Hansmann[l], Solene Patrat-Delon[m], Thierry Prazuck[n], Alice Raffetin[o,p], Pierre Tattevin[q], Gwenaël Vourc'h[b,c], Olivier Lesens[r,s],
Engelbert Mephu Nguifo[a,*]

[a]Université Clermont Auvergne, CNRS, ENSMSE, LIMOS, F-63000 Clermont-Ferrand, France
[b]Université Clermont Auvergne, INRAE, VetAgro Sup, UMR EPIA, 63122 Saint-Genès-Champanelle, France
[c]Université de Lyon, INRAE, VetAgro Sup, UMR EPIA, F-69280 Marcy l'Etoile, France
[d]Infectious and Tropical Diseases Department, CHU Clermont-Ferrand, Clermont-Ferrand, France
[e]CHU Clermont-Ferrand, Inserm, Neuro-Dol, CNRS 6023 Laboratoire Microorganismes: Génome Environnement (LMGE), Université Clermont Auvergne, Clermont-Ferrand, France
[f]Tous Chercheurs Laboratory, UMR 1136 'Interactions Arbres Micro-Organismes', INRAE, Centre INRAE Grand Est-Nancy, F-54280 Champenoux, France
[g]CPIE Champenoux, F-54280 Champenoux, France
[h]INRAE, US 1371 Laboratory of Excellence ARBRE, Centre INRAE Grand Est-Nancy, Champenoux F-54280, France
[i]Infectious Diseases Department, University Hospital of Nancy, Nancy, France
[j]Infectious Disease Department, University Hospital of Saint Etienne, Saint-Etienne, France
[k]IHU-Méditerranée Infection, Marseille, France; Aix Marseille Univ, IRD, AP-HM, SSA, VITROME, Marseille, France
[l]Service des Maladies Infectieuses et Tropicales, Hôpitaux Universitaires, 67000 Strasbourg, France
[m]Infectious Diseases and Intensive Care Unit, Pontchaillou University Hospital, Rennes, France
[n]Department of Infectious and Tropical Diseases, CHR Orléans, Orléans, France
[o]Tick-Borne Diseases Reference Center, North region, Department of Infectious Diseases, Hospital of Villeneuve-Saint-Georges, 40 allée de la Source, 94190 Villeneuve-Saint-Georges
[p]ESGBOR, European Study Group for Lyme Borreliosis
[q]Department of Infectious Diseases and Intensive Care Medicine, Centre Hospitalier Universitaire de Rennes, Rennes, France
[r]Infectious and Tropical Diseases Department, CRIOA, CHU Clermont-Ferrand, Clermont-Ferrand, France
[s]UMR CNRS 6023, Laboratoire Microorganismes: Génome Environnement (LMGE), Université Clermont Auvergne, Clermont-Ferrand, France
[*]Corresponding author
phone: +33473407629, fax: +33473407639, email: engelbert.mephu_nguifo@uca.fr

**Abstract**

Background and objective: Lyme disease which is one of the most common infectious vector-borne diseases manifests itself in most cases with erythema migrans (EM) skin lesions. Recent studies show that convolutional neural networks (CNNs) perform well to identify skin lesions from images. Lightweight CNN based pre-scanner applications for resource-constrained mobile devices can help users with early diagnosis of Lyme disease and prevent the transition to a severe late form thanks to appropriate antibiotic therapy. Also, resource-intensive CNN based robust computer applications can assist non-expert practitioners with an accurate diagnosis. The main objective of this study is to extensively analyze the effectiveness of CNNs for diagnosing Lyme disease from images and to find out the best CNN architectures considering resource constraints.

Methods: First, we created an EM dataset with the help of expert dermatologists from Clermont-Ferrand University Hospital Center of France. Second, we benchmarked this dataset for twenty-three CNN architectures customized from VGG, ResNet, DenseNet, MobileNet, Xception, NASNet, and EfficientNet architectures in terms of predictive performance, computational complexity, and statistical significance. Third, to improve the performance of the CNNs, we used custom transfer learning from ImageNet pre-trained models as well as pre-trained the CNNs with the skin lesion dataset HAM10000. Fourth, for model explainability, we utilized Gradient-weighted Class Activation Mapping to visualize the regions of input that are significant to the CNNs for making predictions. Fifth, we provided guidelines for model selection based on predictive performance and computational complexity.

Results: Customized ResNet50 architecture gave the best classification accuracy of 84.42% ±1.36, AUC of 0.9189±0.0115, precision of 83.1%±2.49, sensitivity of 87.93%±1.47, and specificity of 80.65%±3.59. A lightweight model customized from EfficientNetB0 also performed well with an accuracy of 83.13%±1.2, AUC of 0.9094±0.0129, precision of 82.83%±1.75, sensitivity of 85.21% ±3.91, and specificity of 80.89%±2.95. All the trained models are publicly available at https://dappem.limos.fr/download.html, which can be used by others for transfer learning and building pre-scanners for Lyme disease.

Conclusion: Our study confirmed the effectiveness of even some lightweight CNNs for building Lyme disease pre-scanner mobile applications to assist people with an initial self-assessment and referring them to expert dermatologist for further diagnosis.

**1. Introduction**

Lyme disease is an infectious disease transmitted by ticks and caused by pathogenic bacteria of the *Borrelia burgdorferi* sensu lato group [1]. It is estimated that around 476,000 people in the United States and more than 200,000 people in western Europe are affected by Lyme disease each year [2]. Most of the time an expanding round or oval red skin lesion known as erythema migrans (EM) becomes visible in the victim's body which is the most common early symptom of Lyme disease [1,3]. EM usually appears at the site of a tick bite after one to two weeks (range, 3 to 30 days) as a small redness and expands almost a centimeter per day, creating the characteristic bull's-eye pattern as shown in Figure 1 (a) [1,3–5]. EM generally vanishes within a few weeks or months but the Lyme disease infection advances to affect the nervous system, skin, joints, eyes, and heart [1,4]. Antibiotics can be used as a medium of effective treatment in the early stage of Lyme disease. So, early recognition of EM is extremely important to avoid long-term complications of Lyme disease. Most European and North American guidelines recommend a two-tier serology test to detect antibodies against *Borrelia burgdorferi* sensu lato for diagnosing Lyme disease [6,7]. However, a serology test is only recommended in the absence of EM because early serology has low sensitivity (40% to 60%) and may result in false negatives [6]. Alternatively, direct detection of *Borrelia burgdorferi* sensu lato can be done using culture, microscopy, or PCR [7]. The gold standard of microbiological diagnosis - the culture of bacteria requires laboratory expertise and special media for *Borrelia burgdorferi* sensu lato [6]. Light microscopy-based detection is not feasible in clinical practice [7]. PCR based diagnosis is also very difficult and shows highly variable sensitivity [7]. Direct detection methods are not always feasible for clinicians because of extended processing time and required expertise [8]. The diagnosis of EM is a challenging task because EM can create different patterns instead of the trademark bull's-eye pattern as shown in Figure 1 (b).

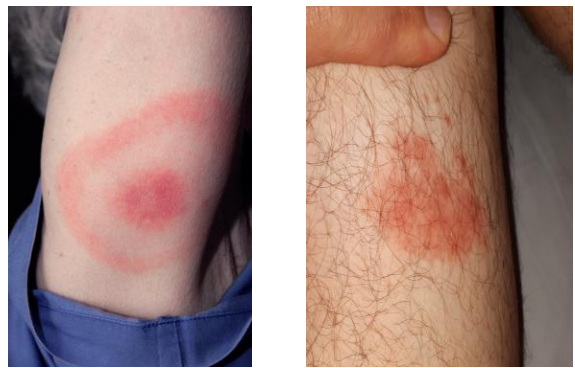

(a) Bull's-eye pattern    (b) Atypical pattern

Figure 1: Patterns of erythema migrans (EM).
(source: https://commons.wikimedia.org/wiki/Category:Erythema_migrans,
Accessed April 1, 2021)

Diagnosing skin disorders requires a careful inspection from dermatologists or infectiologists but their availability, especially in rural areas is scarce [9]. As a result, the diagnosis is generally carried out by non-specialists, and their diagnostic accuracy is in the range of twenty-four to seventy percent [10,11]. The wrong diagnosis can result in improper or delayed treatment which can be harmful to the patient.

Artificial intelligence (AI) powered diagnostic tools can help with the scarcity of expert dermatologists. Recent advancement in deep learning techniques has eased the creation of AI solutions to aid in skin disorder diagnosis. Many works have been done utilizing deep learning techniques specifically convolutional neural networks (CNNs) for diagnosing cancerous and other common skin lesions from dermoscopic images [12–15]. As dermoscopic images require dermatoscopes from dermatology clinics other works have focused on diagnosing skin diseases using deep learning from clinical images [16–18]. According to some of these studies, deep learning-based systems compete on par with expert dermatologists for diagnosing diseases from dermoscopic and clinical images [12–14,17,18].

Despite the vast application of AI in the field of skin lesion diagnosis, there are only a few works related to Lyme disease detection from EM skin lesion images. The unavailability of public EM datasets as a result of privacy concerns of medical data may be the reason for the lack of extensive studies in this field. Čuk et al. [19] proposed a visual system for EM recognition on a private EM dataset using classical machine learning techniques including naïve Bayes, SVM, boosting, and neural nets (not deep learning). They considered ellipse, the common shape of EM and used eccentricity, small and large axis ratio, ellipse angular, and ellipse focus attributes for classification. Deep learning techniques learn image features from training images via an optimization process and recent studies show that image features extracted by deep learning techniques outperform human-engineered image features for medical image classification tasks [8]. Burlina et al. [3] created a dataset of EM by collecting images from the internet and trained a CNN architecture ResNet50 as a binary classifier to distinguish between EM and other skin conditions. Although their dataset is not public, the trained model is publicly available. Burlina et al. [8] further enriched the dataset with more images from the East Coast and Upper Midwest of the United States and trained six CNNs namely ResNet50, InceptionV3, MobileNetV2, DenseNet121, InceptionResNetV2, and ResNet152 for EM classification. Burlina et al. [8] did not make the dataset or the trained models public. Burlina et al. [3] and Burlina et al. [8] used transfer learning from ImageNet [20] pre-trained models and studied the CNNs in terms of predictive performance. With the advancement of CNNs, it is a timely need to extensively study their effectiveness for Lyme disease prediction from EM images.

Lightweight CNN based mobile applications can help people with an initial self-assessment of EM and referring them to expert dermatologist for further diagnosis. Also, resource-intensive CNN based computer applications can assist non-expert practitioners for identifying EM. In this article, our main objective was to study the performance of state-of-the-art CNNs for diagnosing Lyme disease from EM images and to find out the best architecture based on different criteria. As there is no publicly available Lyme dataset of EM images, first, we created a dataset consisting of 866 images of confirmed EM lesions. Images collected from the internet and Clermont-Ferrand University Hospital Center (CF-CHU) of France were carefully labeled into two classes: EM and Confuser, by expert dermatologists and infectiologists from CF-CHU. CF-CHU collected the images from several hospitals in France. Second, we benchmarked twenty-three well-known DCCNs on this dataset in terms of several predictive performance metrics, computational complexity metrics, and statistical significance tests. Best practices for training CNNs on limited data like transfer learning and data augmentation were used. Third, instead of only using transfer learning from models pre-trained on ImageNet dataset we also utilized a dataset of common skin lesions "Human Against Machine with 10000 training images (HAM10000)" [21] for pretraining the CNNs. The use of HAM10000 proved fruitful according to the experimental results. We experimentally searched for the best performing number of layers to unfreeze during transfer learning fine-tuning for each of the studied CNNs. Fourth, for visualizing the regions of the input image that are significant for predictions from the CNN models we used Gradient-weighted Class Activation Mapping (Grad-CAM) [22]. Fifth, we provided guidelines for model selection based on predictive performance and computational complexity. Moreover, we made all the trained models publicly available which can be used for transfer learning and building pre-scanners for Lyme disease. Figure 2 presents the graphical overview of this study.

The rest of the paper is structured as follows: Section 2 contains dataset description, a brief overview of CNN architectures, performance measures, and transfer learning approach used in this study; Section 3 presents experimental studies; Section 4 contains recommendations for model selection, discussion on limitations and scopes; finally, Section 5 presents concluding remarks.

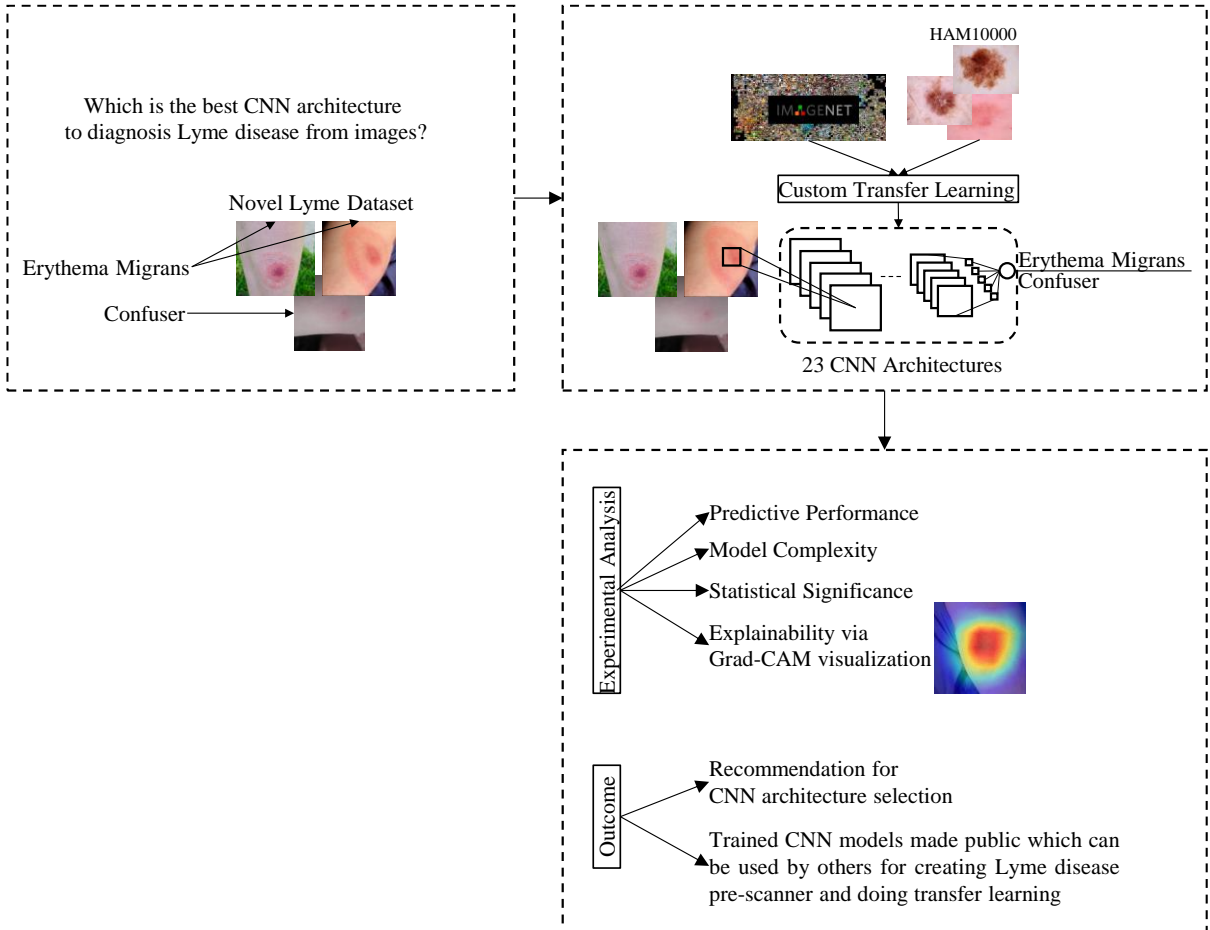

Figure 2: Graphical overview of the study on the effectiveness of CNNs for the diagnosis of Lyme disease from images.

## 2. Methods

The following subsections describe the data organization including data augmentation and cross-validation, a short overview of the considered CNN architectures, performance measures, explainability method, and the transfer learning approach used in this study.

### 2.1. Dataset Preparation

As a labeled public dataset is not available for Lyme disease prediction from EM images, we created a dataset by collecting images from the internet and CF-CHU. CF-CHU collected EM images from several hospitals located in France. The use of images from the internet was inspired by related skin lesion analysis studies [3,8,17]. Duplicate images were removed using an image hashing-based duplicate image detector followed by the removal of inappropriate images through human inspection. After the initial curation steps, we got a total of 1672 images. Expert dermatologists and infectiologists from CF-CHU classified the curated images into two categories: EM and Confuser, making it a two-class classification problem. Out of 1672 images, 866 images were assigned to EM class and 806 images were assigned to Confuser class.

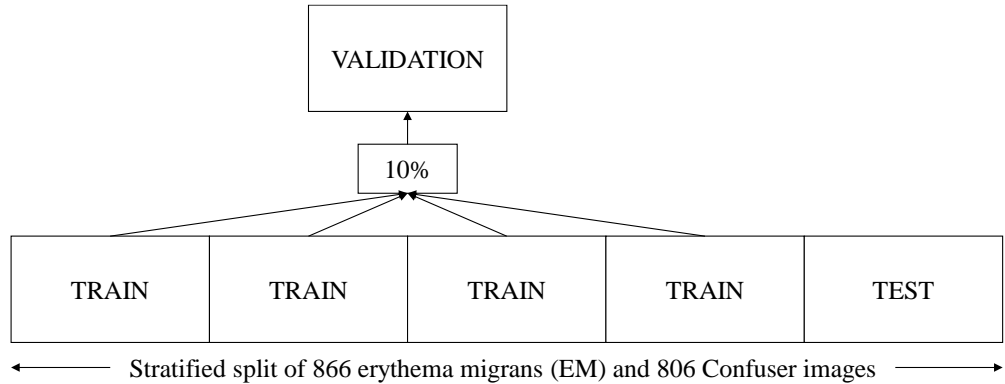

Figure 3: Five-fold cross-validation setup.

We further subdivided the dataset into five-folds using stratified five-fold cross-validation to make sure each of the folds maintains the original class ratio. One of the folds was used as a test set and the remaining four were used as the training set with a rotation of the folds for five runs. Each time, 10% of the training data was assigned to the validation set as shown in Figure 3.

DCCNs require a considerable amount of data for training and data augmentation can help with expanding the dataset. We applied data augmentation techniques only to the training set to expand it twenty times. We used data augmentation to create 20 images from a single image both for EM and Confuser categories because our dataset was balanced. We used flip (vertical or horizontal), rotation, brightness, contrast, and saturation augmentation by considering the best performing augmentations for skin lesions [15]. Besides, we also used perspective skew transformation to cover the case of looking at a picture from different angles. Augmentor [23] an image augmentation library specially built for biomedical image augmentation was used for applying the augmentations. We used 0.5 as the probability of applying each of the augmentation operations. Rotation operation was performed with a maximum rotation angle of 5 degrees. We also used random rotation by either 90, 180, or 270 degrees. Brightness, contrast, and saturation augmentations were performed with a minimum adjustment factor of 0.7 and a maximum adjustment factor of 1.3. For all the other parameters we used default values provided by Augmentor library. Figure 4 shows some example images resulting from augmentations applied on a sample image.

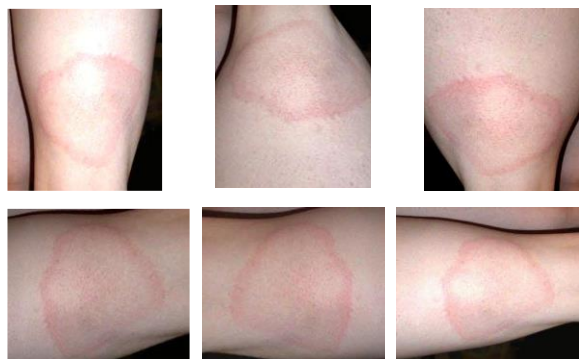

Figure 4: Data augmentation examples.

## 2.2. Brief Overview of CNN Architectures for Lyme Disease Diagnosis

CNNs are a kind of neural network that simulates some actions generated in the human visual cortex using convolution mathematical operation to extract features from input and passing these features through successive layers generates more abstract features to yield a final output [24]. CNNs are modular in design, where convolution-based building blocks are repeatedly stacked for feature extraction with pooling layers placed in between for reducing feature space, learnable parameters, and controlling overfitting [25]. Starting with LeNet [24] in 1988 the popularity of CNNs increased with AlexNet [26] winning the ImageNet Large Scale Visual Recognition Challenge (ILSVRC) [20] of 2012. As a result of the effectiveness of CNNs in solving complex problems, several CNN architectures have been introduced over the past few years. The following subsections provide a brief overview of the CNN architectures used in this study for diagnosing Lyme disease from EM images.

### 2.2.1. VGG Architecture

VGG architecture [27] is based on the idea of deeper networks with smaller filters ($3 \times 3$). There are thirteen convolutional layers and three fully connected layers in VGG16 architecture as shown in Figure 5. Another variation of VGG architecture called VGG19 has sixteen convolutional layers and three fully connected layers. VGG architecture showed better effectiveness of deeper architectures in terms of predictive performance but requires training a huge number of parameters. To the best of our knowledge, VGG architectures have not been used previously for Lyme disease analysis.

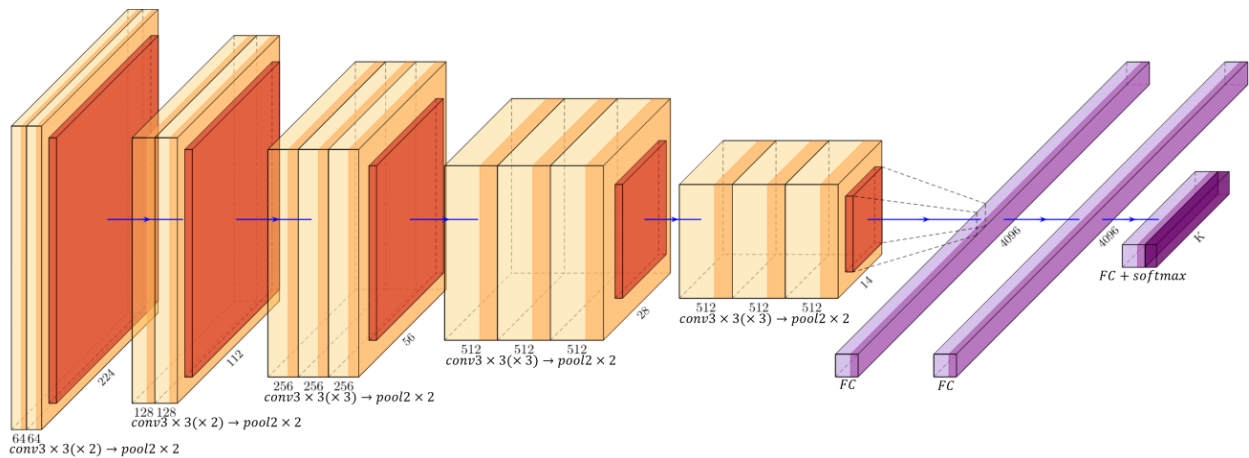

Figure 5: VGG16 architecture. Input image is of shape $224 \times 224 \times 3$. $FC$ stands for fully connected layer and $K$ is the number of target classes.

### 2.2.2. Inception Architecture

Inception architecture [28] uses inception module as shown in Figure 6, which is a combination of several convolution layers with small filters ($1 \times 1, 3 \times 3, 5 \times 5$) applied simultaneously on the same input to facilitate the extraction of more information. The output filter banks from the convolution layers of inception module are concatenated into a single vector, which is served as the input for next stage. To reduce learnable parameters and computational complexity inception module uses $1 \times 1$ convolution at the beginning of convolution layers. InceptionV1 architecture is the winner of ILSVRC 2014 competition, and it's also known as GoogleNet. Further improvement resulted in the creation of several versions of inception architectures named InceptionV2, InceptionV3, and InceptionV4 [29,30]. InceptionV2 and InceptionV3 improved the architecture with smart factorized convolution, batch normalized auxiliary classifier, and label smoothing whereas, InceptionV4 focused on the uniformity of the architecture with more inception modules than InceptionV3. Burlina et al. [8] used ImageNet pre-trained InceptionV3 architecture for Lyme disease analysis.

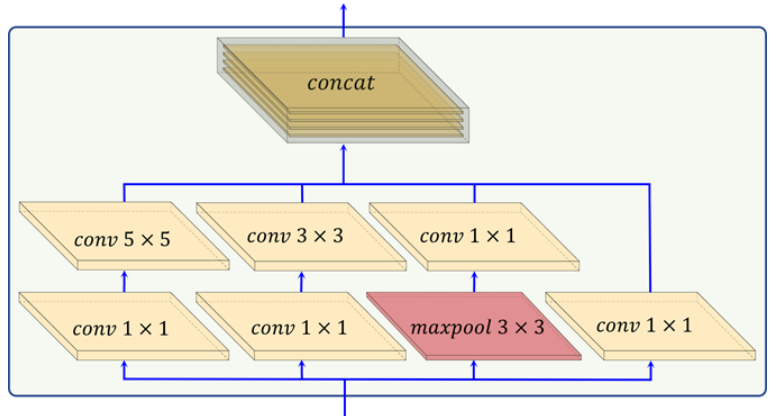

Figure 6: Inception module of Inception architecture.

### 2.2.3. ResNet Architecture

ResNet architecture [31] tried to solve the vanishing gradient and accuracy degradation problems of deep models by introducing residual block with identity shortcut connection that directly connects the input to the output of the block allowing the gradient to flow through the shortcut path as shown in Figure 7. It's the winner of ILSVRC 2015 competition. Depending on the number of weight layers there are many variants of ResNet architecture such as ResNet18, ResNet34, ResNet50, ResNet101, ResNet152, ResNet164, ResNet1202, etc., where the number represents the count of weight layers. He et al. [32] proposed ResNetV2 with pre-activation of the weight layers as opposed to the post-activation of original ResNet architecture. InceptionResNet is a hybrid of Inception and ResNet architecture having two variations named InceptionResNetV1 and InceptionResNetV2, which differ mainly in terms of the number of used filters [30]. Burlina et al. [3] used ImageNet pre-trained ResNet50 and Burlina et al. [8] used ImageNet pre-trained ResNet50, ResNet152, and InceptionResNetV2 architectures for Lyme disease analysis.

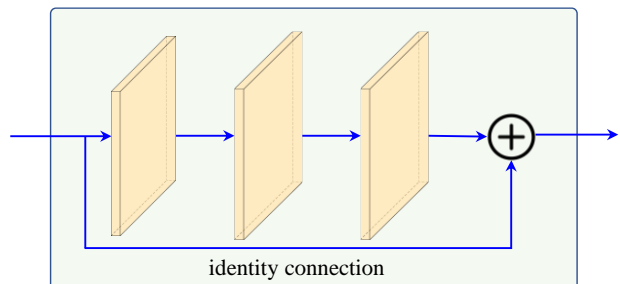

Figure 7: Residual block of ResNet architecture.

### 2.2.4. DenseNet Architecture

Dense Convolutional Network (DenseNet) [33] extended ResNet by introducing dense blocks where each layer within a dense block receives inputs from all the previous layers as shown in Figure 8. DenseNet concatenates the incoming feature maps of a layer with output feature maps instead of summing them up as done in ResNet. Dense blocks within DenseNet are connected with transition layers consisting of convolution and pooling to perform the required downsampling operation. Depending on the number of weight layers there are several versions of DenseNet like DenseNet121, DenseNet169, DenseNet201, DenseNet264, etc. Besides solving the vanishing gradient problem DenseNet also eases feature propagation and reuse, and a reduction in the number of learnable parameters compared to ResNet. Burlina et al. [8] used ImageNet pre-trained DenseNet121 architecture for Lyme disease analysis.

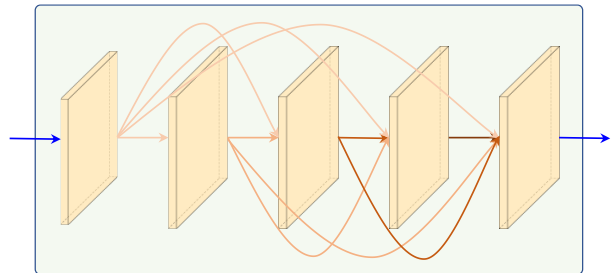

Figure 8: Building block of DenseNet architecture.

### 2.2.5. MobileNet Architecture

MobileNetV1 [34] used depthwise separable convolution extensively to reduce the computational cost. Standard convolution performs spatial and channel-wise computations one step but depthwise separable convolution first applies separate convolutional filter for each input channel and then uses pointwise convolution on concatenated channels to produce required number of output channels as shown in Figure 9. MobileNetV1 was designed to run very efficiently on mobile and embedded devices. MobileNetV2 [35] improved upon the concepts of MobileNetV1 by incorporating thin linear bottlenecks with shortcut connections between the bottlenecks as shown in Figure 10. This is called inverted residual block as it uses narrow → wide → narrow as opposed to the wide → narrow → wide architecture of traditional residual block. MobileNetV3 [36] incorporated squeeze-and-excitation layers [37] in the building block of MobileNetV2 which provides channel-wise attention and used MnasNet [38] to search for a coarse architecture that was further optimized with NetAdapt [39] algorithm. Burlina et al. [8] used ImageNet pre-trained MobileNetV2 architecture for Lyme disease analysis.

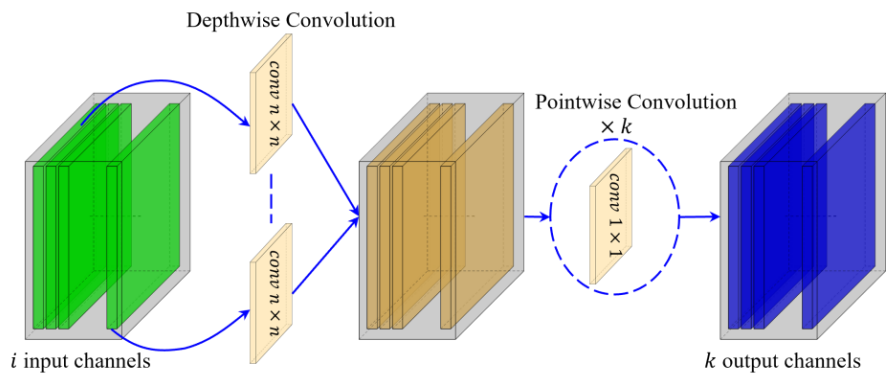

Figure 9: Depthwise separable convolution.

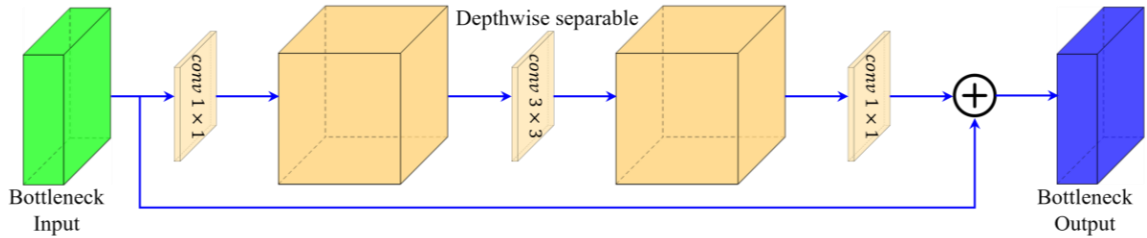

Figure 10: Building block of MobileNetV2 architecture.

### 2.2.6. Xception architecture

Extreme version of Inception the Xception architecture [40] replaced the Inception module with a modified version of depthwise separable convolution where the order of depthwise convolution and pointwise convolutions are reversed as shown in Figure 11. Xception also uses shortcut connections like ResNet architecture. On ImageNet dataset Xception performs slightly better than the InceptionV3 architecture. To the best of our knowledge, Xception architecture has not been used previously for Lyme disease analysis.

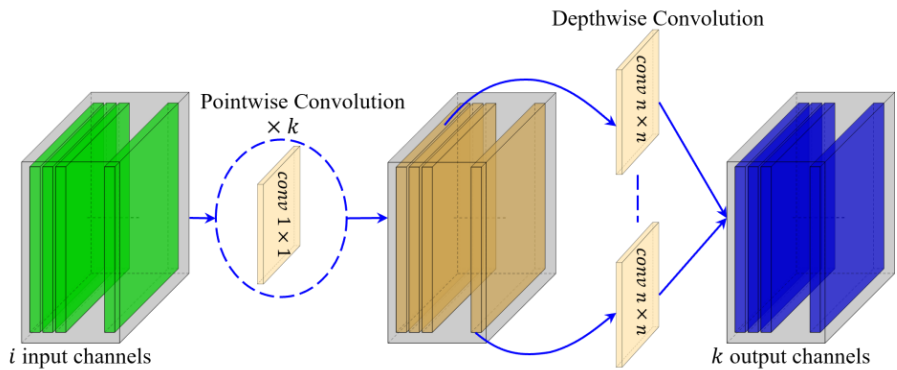

Figure 11: Building block of Xception architecture.

### 2.2.7. NASNet architecture

Neural Architecture Search Netowork [41] from Google Brain utilizes reinforcement learning with a Recurrent Neural Network based controller to search for efficient building block for a smaller dataset which is then transferred to a larger dataset by stacking multiple copies of the found building block. NASNet blocks are comprised of normal and reduction cells as shown in Figure 12. Normal cells produce feature map of same size as input whereas reduction cells reduce the size by a factor of two. NASNet optimized for mobile applications is called NASNetMobile whereas the larger version is called NASNetLarge. To the best of our knowledge, NASNet architectures have not been used previously for Lyme disease analysis.

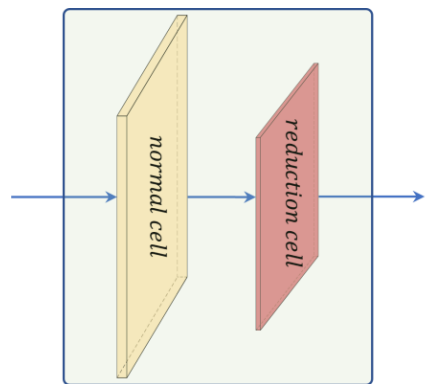

Figure 12: Building block of NASNet architecture.

### 2.2.8. EfficientNet architecture

EfficientNet [42] which is among the most efficient models proposed a scaling method to uniformly scale all dimensions of depth, width, and resolution of a network using a compound coefficient. The baseline network of EfficientNet was built with neural architecture search incorporating squeeze-and-excitation in the building block of MobileNetV2. The scaling method is defined as:

$$depth = \alpha^{\emptyset}$$
$$width = \beta^{\emptyset}$$
$$resolution = \gamma^{\emptyset},$$
$$\text{s.t. } \alpha.\beta^2.\gamma^2 \approx 2; \alpha \geq 1, \beta \geq 1, \gamma \geq 1 \quad (1)$$

where, the coefficient $\emptyset$ controls available resources and $\alpha, \beta,$ and $\gamma$ are constants obtained by grid search. EfficientNetB0-B7 are a family of architectures scaled up from the baseline network that reflects a good balance of accuracy and efficiency. To the best of our knowledge, EfficientNet architectures have not been used previously for Lyme disease analysis.

### 2.3. Predictive Performance Measures

To compare the predictive performance of the trained CNN models we used accuracy, recall/sensitivity/hit rate/true positive rate (TPR), specificity/selectivity/true negative rate (TNR), precision/ positive predictive value (PPV), negative predictive value (NPV), Matthews correlation coefficient (MCC), Cohen's kappa coefficient ($\kappa$), positive likelihood ratio (LR+), negative likelihood ratio (LR-), F1-score, confusion matrix and area under the receiver operating characteristic (ROC) curve (AUC) metrics. Confusion matrix is a way of presenting the count of true negatives (TN), false positives (FP), false negatives (FN), and true positives (TP) in a matrix format where the y-axis presents true labels and x-axis presents predicted labels. Accuracy measures the proportion of correctly classified predictions among all the predictions, and it is calculated as:

$$Accuracy = \frac{TP + TN}{TP + TN + FP + FN} \quad (2)$$

Recall/sensitivity/hit rate/TPR measures the proportion of actual positives correctly identified, and it is expressed as:

$$Recall, Sensitivity, hit\ rate, TPR = \frac{TP}{TP + FN} \quad (3)$$

Specificity/selectivity/ TNR measures the proportion of actual negatives correctly identified, and it is expressed as:

$$Specificity, Selectivity, TNR = \frac{TN}{TN + FP} \quad (4)$$

Precision/ PPV measures the proportion of correct positive predictions, and it is calculated as:

$$Precision,\ PPV = \frac{TP}{TP + FP} \tag{5}$$

NPV measures the proportion of negative predictions that are correct, and it is calculated as:

$$NPV = \frac{TN}{TN + FN} \tag{6}$$

MCC provides a summary of the confusion matrix, and it is calculated as:

$$MCC = \frac{TP * TN - FP * FN}{\sqrt{(TP + FP)(TP + FN)(TN + FP)(TN + FN)}} \tag{7}$$

MCC value is in the range $[-1, +1]$, where 0 is like random prediction, $+1$ means a perfect prediction, and $-1$ represents inverse prediction. Cohen's kappa coefficient ($\kappa$) metric is used to assess inter-rater agreement which tells us how the model is performing compared to a random classifier, and it is calculated with the formula:

$$\kappa = \frac{p_o - p_e}{1 - p_e} \tag{8}$$

where, $p_o$ is the relative observed agreement among the raters and $p_e$ is the hypothetical probability of expected agreement which is defined for $c$ categories as:

$$p_e = \frac{1}{N^2} \sum_c n_{c1} n_{c2} \tag{9}$$

where, $N$ is the total number of observations, and $n_{cr}$ is the number of predictions of category $c$ by rater $r$. Value of $\kappa$ is in the range $[-1, +1]$, where a value of 1 indicates perfect agreement, 0 means agreement only by chance, and a negative value indicates the agreement is worse than the agreement by chance. Likelihood ratio (LR) is used for assessing the potential utility of performing a diagnostic test and it is calculated for both positive test and negative test results called LR+ and LR-, respectively. LR+ is the ratio of the probability of a person having a disease testing positive to the probability of a person without the disease testing positive. LR- is the ratio of the probability of a person having the disease testing negative to the probability of without the disease testing negative. LR+ and LR- are calculated based on sensitivity and specificity values using the following formulas:

$$LR+ = \frac{sensitivity}{1 - specificity} \tag{10}$$

$$LR- = \frac{1 - sensitivity}{specificity} \tag{11}$$

A value of LR greater than 1 shows increased evidence. F1-Score combines precision and recall, and it is defined as the harmonic mean of precision and recall as follows:

$$F1 - Score = 2 * \frac{Precision * Recall}{Precision + Recall} \tag{12}$$

ROC curve is a plot of TPR against false positive rate (FPR) at various threshold settings where FPR is defined as:

$$FPR = \frac{FP}{FP + TN} \tag{13}$$

Area under the ROC curve (AUC) is the measure of the classifier's ability to separate between classes and the higher the AUC, the better the ability of the classifier for separating the positive class from the negative class.

As our dataset was balanced so, accuracy can be considered a good measure of predictive performance [43] and we did most of the analysis in terms of accuracy but also kept the other metrics to provide insights for experts from different domains as done in relevant studies [3,8]. We also used critical difference (CD) diagram [44] to rank the CNN models in terms of accuracy and to show the statistically significant difference in predictive performance. A thick horizontal line connects a group of models in the CD diagram that are not significantly different in terms of predictive performance. We used non-parametric Friedman test [45] to reject the null hypothesis of statistical similarity among all the models followed by Nemenyi post-hoc all-pair comparison test [46] for showing the difference among the models at a significance level, $\alpha = 0.1$. We kept the confusion matrix, ROC curve, and cross-validation fold-wise details of all the trained models at https://dappem.limos.fr/sdata.html to keep the paper concise and readable.

### 2.4. Model Complexity Measures

To compare the trained CNNs in terms of complexity we used the total number of model parameters, the total number of floating-point operations (FLOPs), average training time per epoch, disk and GPU memory usage, and average inference time per image. FLOPs reveal how computationally costly a model is and we counted FLOPs for each of the models using TensorFlow profiler [47] considering a batch size of one. For reporting the average training time per epoch, we calculated the average of the training time of three epochs during transfer learning fine-tuning. Disk usage of a CNN model is the amount of storage required to save the model architecture along with weights. We calculated the GPU usage of a CNN model by inspecting the memory allocated in GPU after loading a trained instance of the model. To measure the average inference time per image of a model we took the average of three hundred inferences on the same input image.

### 2.5. Model Explainability

Explainability is important for AI tools especially in the case of medical applications [48]. We used Grad-CAM explainability technique for visualizing the regions of the input image that are significant for predictions from the CNN models as shown in Figure 13. Grad-CAM uses gradient flowing into the ultimate convolution layer for producing heatmaps, and it is a kind of post-hoc attention that can be applied on an already trained model. Grad-CAM provides similar result to occlusion sensitivity map [49] that works by masking patches of the input image, but Grad-CAM is much faster to calculate compared to image occlusion [22].

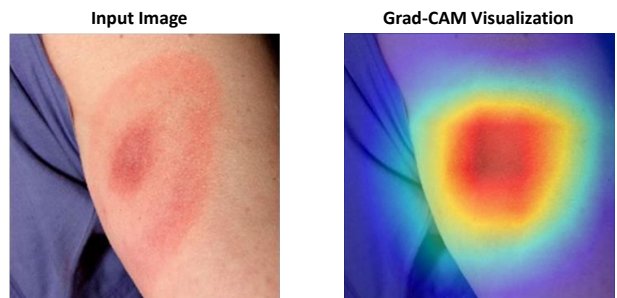

Figure 13: Gard-CAM visualization example.

## 2.6. Transfer Learning Approach

In this study, we used transfer learning as our Lyme dataset is not huge enough to obtain good performance by training large CNNs from scratch. We started with a CNN already pre-trained on ImageNet dataset and after removing the original ImageNet classification head our EM classification head consisting of Global Average Pooling (GAP) layer, dropout layer, and a fully connected layer with sigmoid activation for binary classification was added as shown in Figure 14. According to our experiments fine-tuning the whole CNN architecture after training the classifier head with our Lyme dataset performed poorly compared to the partial fine-tuning of several layers at the end of the CNN while keeping rest of the layers frozen. We empirically found out the number of layers $U$ to fine-tune from $N$ number of ImageNet pre-trained layers for each of the CNN architectures used in this study. According to the experimental results pretraining the unfrozen part with HAM10000 dataset further improved the performance of the DCCNs.

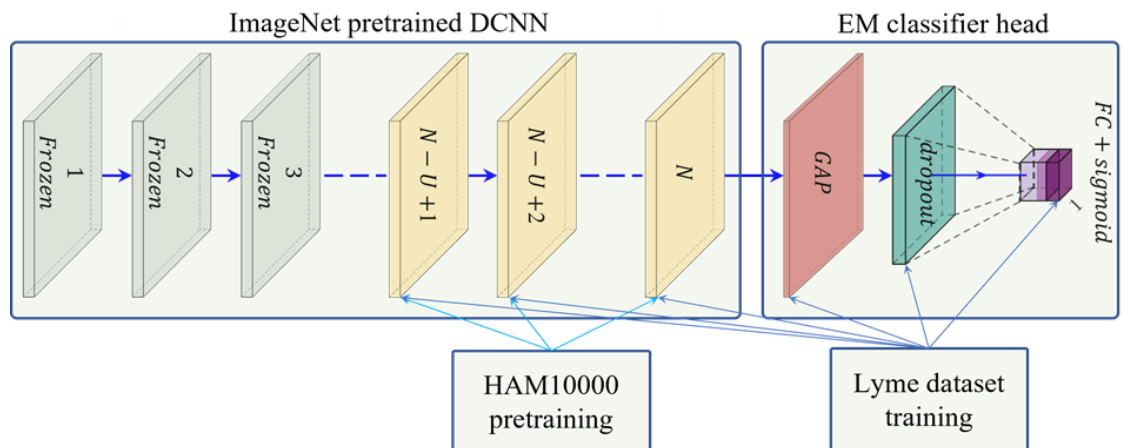

Figure 14: Transfer learning workflow used in this study. $GAP$ stands for Global Average Polling. $N$ is the number of ImageNet pre-trained layers and $U$ represents the number of layers used for fine-tuning.

## 3. Experimental Studies

The following subsections describe experimental settings including model selection and parameter settings, software and hardware used for the study, the experimental results, and recommendations for model selection.

## 3.1. Experimental Settings

In this study, we benchmarked twenty-three CNN models, namely VGG16, VGG19, ResNet50, ResNet101, ResNet50V2, ResNet101V2, InceptionV3, InceptionV4, InceptionResNetV2, Xception, DenseNet121, DenseNet169, DenseNet201, MobileNetV2, MobileNetV3Large, MobileNetV3Small, NASNetMobile, EfficientNetB0, EfficientNetB1, EfficientNetB2, EfficientNetB3, EfficientNetB4, and EfficientNetB5 for diagnosing Lyme disease from EM images. These models were selected to explore a diverse set of CNN models covering various prospects, like different architectures, depths, and complexities.

To the best of our knowledge ResNet50 is the only publicly available trained CNN that was used for Lyme disease identification by Burlina et al. [3]. We are calling this model ResNet50-Burlina which is a collection of five models (trained on five-fold cross-validation data) available at https://github.com/neil454/lyme-1600-model. We did extensive analysis on the ResNet50 architecture using seven different transfer learning configurations: (i) training ResNet50 model on our Lyme dataset from scratch without using transfer learning (called ResNet50-NTL, where, **NTL** stands for no transfer learning), (ii) pretraining ResNet50 model with only HAM10000 data followed by fine-tuning all the layers with our Lyme dataset (called ResNet50-HAM-FFT, where **HAM** means HAM10000 and **FFT** stands for full fine-tuning), (iii) training only the EM classifier head of an ImageNet pre-trained ResNet50 model with our Lyme dataset (called ResNet50-IMG-WFT, where **IMG** means ImageNet and **WFT** stands for without fine-tuning), (iv) fine-tuning all the layers of ImageNet pre-trained ResNet50 model with our Lyme dataset (called ResNet50-IMG-FFT), (v) fine-tuning $U$ no of layers of an ImageNet pre-trained ResNet50 model with our Lyme dataset (called ResNet50-IMG-FT$U$, where **FT$U$** means fine-tuning $U$ no of layers), (vi) pretraining the whole ImgaeNet pre-trained ResNet50 model by HAM10000 data before fine-tuning $U$ layers with our Lyme dataset (called ResNet50-IMG-HAMFP-FT$U$, where, **HAMFP** means full pre-training with HAM10000 dataset), and (vii) pretraining only the unfrozen $U$ layers of a ImgaeNet pre-trained ResNet50 model with HAM10000 data before fine-tuning $U$ layers with our Lyme dataset (called ResNet50-IMG-HAMPP-FT$U$, where, **HAMPP** means partial pre-training with HAM10000 dataset). To see the effect of data augmentation, we trained a ResNet50 model without data augmentation and transfer learning (called ResNet50-NoAug, where NoAug means no data augmentation). All the other models were trained with data augmentation as described in section 2.1. According to experimental results (discussed below), ResNet50-IMG-HAMPP-FT$U$ configuration performed best, and we used this configuration (pretraining only the unfrozen $U$ layers of an ImgaeNet pre-trained model with HAM10000 data before fine-tuning $U$ layers with our Lyme dataset) for training all the architectures used in this study. For simplicity, the best performing trained models of each of the architectures are presented in ModelName-$U$ format, where $U$ represents the no of unfrozen layers. For example, EfficientNetB0-187 means EfficientNetB0-IMG-HAMPP-FT187 and ResNet50-141 means ResNet50-IMG-HAMPP-FT141.

For training all the models, we used a dropout rate of 0.2 for the dropout layer in EM classifier head section. Adaptive Moment Estimation (ADAM) [50] optimizer with exponential decay rate for the first and second moment estimates set to 0.9 and 0.999 respectively was used with a learning rate of 0.0001 for training the classifier head and 0.00001 for fine-tuning. We also used early stopping to terminate the training if there was no improvement in validation accuracy for ten epochs. A batch size of 32 was used. For reporting the number of layers to unfreeze during transfer learning, we stated the total number of layers to unfreeze including layers containing both trainable and non-trainable parameters.

We used three NVIDIA QUADRO RTX 8000 GPUs and two Desktop Computers with Intel Xeon W-2175 processor, 64 GB DDR4 RAM, and Ubuntu 18.04 operating system to perform all the experiments. Python v3.6.9, and TensorFlow v2.4.1 platform [47] were used for all the implementations and experimentations of this study.

### 3.2. Experimental Results

Čuk et al. [19] reported accuracies in the range of 69.23% to 80.42% using classical machine learning methods whereas, Burlina et al. [8] reported the best accuracy of 81.51% using ResNet50 architecture for the case of EM vs all classification problems. There was a common subset of images collected from the internet in both the dataset of Burlina et al. [3] and our Lyme dataset. ResNet50-Burlina model gave an accuracy of 76.05% when tested on our full dataset as shown in Table 1.

Table 1: Performance metrics of ResNet50-Burlina model trained by Burlina et al. [3] tested on the whole dataset of this study. Within each cell, the value after (±) symbol represents the standard deviation.

| Metric / Model | Accuracy | Sensitivity | Specificity | Precision | NPV | MCC | Kappa | LR+ | LR- | F1-Score | AUC |
|---|---|---|---|---|---|---|---|---|---|---|---|
| ResNet50-Burlina | 76.05 ±0.74 | 70.05 ±3.6 | 82.51 ±3.31 | 81.29 ±2.1 | 72.04 ±1.71 | 0.5294 ±0.0132 | 0.5229 ±0.0145 | 4.1017 ±0.5172 | 0.362 ±0.0309 | 0.7515 ±0.0137 | 0.481 ±0.0509 |

Table 2 presents the predictive performance measures of our experimentation with ResNet50 architecture. ResNet50-NoAug model resulting from training a ResNet50 architecture from scratch without using data augmentation and transfer learning gave an accuracy of 61.42%. ResNet50-NTL model obtained by training ResNet50 architecture with data augmentation and without transfer learning improved the accuracy to 76.35%. So, data augmentation provided large gain in predictive performance (ResNet50-NTL compared to ResNet50-NoAug). ResNet50-HAM-FFT model resulting from pretraining ResNet50 architecture with only HAM10000 data followed by fine-tuning of all the layers with our Lyme dataset showed a degraded accuracy of 72.27%. ResNet50-IMG-WFT, generated by training only the EM classifier head of an ImageNet pre-trained ResNet50 architecture improved the accuracy to 78.94%. ResNet50-IMG-FFT, resulting from fine-tuning all the layers of ImageNet pre-trained ResNet50 architecture, further improved the classification accuracy to 82.22%. Whereas ResNet50-IMG-FT141, model resulting from fine-tuning 141 layers of pre-trained ResNet50 architecture gave an accuracy of 83.24% which is better compared to unfreezing the full architecture. ResNet50-IMG-HAMFP-FT141, model resulting from pretraining the whole ImgaeNet pre-trained ResNet50 model by HAM10000 data before fine-tuning 141 layers with our Lyme dataset reduced the accuracy to 82.35%. But pretraining only the unfrozen 141 layers with HAM10000 data gave us the model ResNet50-IMG-HAMPP-FT141 with the best accuracy of 84.42%. Figure 15 shows the CD diagram in terms of accuracy for these ResNet50 based models. We excluded ResNet50-Burlina model from this diagram because the model was tested on the whole dataset as opposed to other configurations. The Friedman test null hypothesis was rejected with a *p* value of 0.00003. From the CD diagram, we can see that ResNet50-IMG-HAMPP-FT141 achieved the best average ranking among the compared models. Although there is no statistically significant difference among ResNet50-IMG-FFT, ResNet50-IMG-FT141, ResNet50-IMG-HAMFP-FT141, and ResNet50-IMG-HAMPP-FT141 in terms of accuracy the ResNet50-IMG-HAMPP-FT141 model performed better in terms of most of the metrics (7 out of 11) as highlighted in Table 2. To summarize, pretraining only the unfrozen part of an ImageNet pre-trained CNN with HAM10000 data provided the best accuracy according to our experiments. So, for all the other CNN architectures, we only reported the performance resulting from this configuration.

Performance metrics for the best performing configuration of all the CNN architectures used in this study are shown in Table 3. All these models used HAM10000 pretraining for the unfrozen part of the network. The number at the end of the model's name represents the number of layers unfrozen during transfer learning fine-tuning. ResNet50-141 achieved the best accuracy of 84.42%. Most of the models except MobileNetV2-62, MobileNetV3Small-182, and NASNetMobile-617 showed good AUC values of above 90% and good sensitivity suggesting that these CNNs can be a good choice for building pre-scanners for Lyme disease. Figure 16 shows the CD diagram in terms of accuracy for these models. The Friedman test null hypothesis was rejected with a $p$ value of 0.0822. From the CD diagram, we can see that ResNet50-141 achieved the best average ranking followed by VGG19-13 and DenseNet121-379 respectively. Xception and Inception-based architectures had a similar ranking. NasNetMobile-617 ranked worst among all the models. The accuracy of the models varied from 81.3% to 84.42% and there is no statistically significant difference in terms of accuracy metric among most of the trained models. Overall, ResNet50-141 performed better in terms of various metrics (5 out of 11) as highlighted in Table 3.

Table 2: Five-fold cross-validation performance metrics of ResNet50 based models. Within each cell, the value after (±) symbol represents the standard deviation across five folds. Bold indicates the best result for each of the metrics.

| Metric / Model | Accuracy | Sensitivity | Specificity | Precision | NPV | MCC | Kappa | LR+ | LR- | F1-Score | AUC |
|---|---|---|---|---|---|---|---|---|---|---|---|
| ResNet50-NoAug | 61.42 ±1.29 | 71.73 ±8.65 | 50.37± 8.79 | 61.0 ±1.5 | 63.03 ±3.17 | 0.2302 ±0.0234 | 0.2224 ±0.0256 | 1.4592 ±0.0863 | 0.5497 ±0.0764 | 0.656 ±0.0325 | 0.6505 ±0.0216 |
| ResNet50-NTL | 76.35 ±2.43 | 78.49 ±8.47 | 74.04 ±4.6 | 76.64 ±1.64 | 76.92 ±5.22 | 0.5305 ±0.0431 | 0.5261 ±0.0464 | 3.0735 ±0.2867 | 0.2853 ±0.0906 | 0.7723 ±0.0398 | 0.8471 ±0.0185 |
| ResNet50-HAM-FFT | 72.27 ±1.69 | 75.85 ±1.27 | 68.42 ±4.05 | 72.18 ±2.55 | 72.48 ±1.08 | 0.4447 ±0.0341 | 0.4435 ±0.0347 | 2.4434 ±0.3248 | 0.3536 ±0.0193 | 0.7393 ±0.0116 | 0.7979 ±0.0251 |
| ResNet50-IMG-WFT | 78.94 ±1.48 | 82.55 ±2.77 | 75.06 ±5.11 | 78.27 ±3.2 | 80.11 ±1.77 | 0.5799 ±0.03 | 0.5772 ±0.0305 | 3.4636 ±0.7671 | 0.2316 ±0.0255 | 0.8025 ±0.0101 | 0.8666 ±0.0163 |
| ResNet50-IMG-FFT | 82.22 ±1.36 | 85.27 ±2.67 | 78.93 ±5.26 | 81.55 ±3.42 | 83.42 ±1.63 | 0.6458 ±0.0262 | 0.6431 ±0.028 | 4.3127 ±1.0994 | 0.1854 ±0.0226 | 0.8326 ±0.0083 | 0.909 ±0.0092 |
| ResNet50-IMG-FT141 | 83.24 ±1.04 | 85.29 ±2.27 | **81.04** ±2.28 | 82.91 ±1.49 | 83.74 ±1.96 | 0.6649 ±0.0212 | 0.6641 ±0.021 | 4.5575 ±0.493 | 0.1812 ±0.0255 | 0.8405 ±0.0104 | 0.9134 ±0.0091 |
| ResNet50-IMG-HAMFP-FT141 | 82.35 ±1.62 | **89.28** ±2.42 | 74.91 ±5.11 | 79.45 ±3.05 | **86.81** ±2.03 | 0.6521 ±0.0295 | 0.6448 ±0.0333 | 3.7072 ±0.7368 | **0.1421** ±0.0251 | 0.84 ±0.0111 | 0.9113 ±0.0091 |
| ResNet50-IMG-HAMPP-FT141 | **84.42** ±1.36 | 87.93 ±1.47 | 80.65 ±3.59 | **83.1** ±2.49 | 86.19 ±1.27 | **0.6893** ±0.0263 | **0.6874** ±0.0277 | **4.703** ±0.8624 | 0.1493 ±0.0155 | **0.8541** ±0.0106 | **0.9189** ±0.0115 |

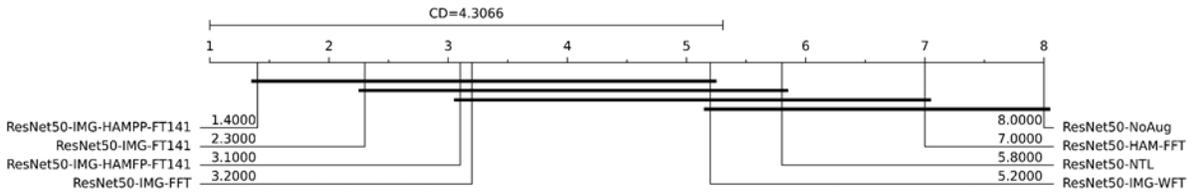

Figure 15: Accuracy critical difference diagram for ResNet50 models. The models are ordered by best to worst average ranking from left to right. The number beside a model's name represents the average rank of the model. CD is the critical difference for Nemenyi post-hoc test. Thick horizontal line connects the models that are not statistically significantly different.

Table 4 summarizes the complexities of the CNN models used in this study. MobileNetV3Small-182 was the most lightweight model with the lowest number of parameters, FLOPs, and memory usage. InceptionResNetV2-500 has the highest number of parameters and memory usage and slowest inference time. Xception-118 was the fastest in terms of inference time. VGG19-13 required the highest number of FLOPs. ResNet50V2-105 required the least amount of time to train on average whereas, EfficientNetB5-444 was the slowest to train.

Figure 17 shows the Grad-CAM visualizations of the models trained on the same training fold for two test images. The test images are shown at the lower right corner of the image. From the figure, it can be seen that different versions of EfficientNet focused more on the lesion part of the image compared to other models.

Table 3: Five-fold cross-validation performance metrics for the best performing configurations of the trained CNN models. Within each cell, the value after (±) symbol represents the standard deviation across five folds. Bold indicates the best result for each of the metrics.

| Metric / Model | Accuracy | Sensitivity | Specificity | Precision | NPV | MCC | Kappa | LR+ | LR- | F1-Score | AUC |
|---|---|---|---|---|---|---|---|---|---|---|---|
| VGG16-8 | 82.17 ±1.23 | 85.77 ±3.58 | 78.31 ±4.36 | 81.12 ±2.62 | 83.88 ±3.02 | 0.6453 ±0.0253 | 0.6422 ±0.0249 | 4.0983 ±0.7329 | 0.1802 ±0.0388 | 0.8328 ±0.0116 | 0.9011 ±0.0079 |
| VGG19-13 | 84.14 ±1.62 | 85.29 ±1.69 | 82.9 ±2.63 | **84.32** ±1.97 | 84.0 ±1.67 | 0.6826 ±0.0323 | 0.6823 ±0.0326 | 5.0924 ±0.6884 | 0.1777 ±0.0214 | 0.8479 ±0.0146 | 0.913 ±0.0074 |
| ResNet50-141 | **84.42** ±1.36 | 87.93 ±1.47 | 80.65 ±3.59 | 83.1 ±2.49 | 86.19 ±1.27 | **0.6893** ±0.0263 | **0.6874** ±0.0277 | 4.703 ±0.8624 | 0.1493 ±0.0155 | **0.8541** ±0.0106 | **0.9189** ±0.0115 |
| ResNet101-150 | 82.64 ±2.1 | 83.68 ±3.49 | 81.52 ±2.29 | 82.97 ±1.8 | 82.4 ±3.09 | 0.6528 ±0.0419 | 0.6522 ±0.0418 | 4.6001 ±0.6257 | 0.2004 ±0.0427 | 0.8329 ±0.022 | 0.9044 ±0.0094 |
| ResNet50V2-105 | 82.37 ±2.15 | 85.53 ±3.35 | 78.96 ±6.13 | 81.66 ±3.83 | 83.72 ±2.63 | 0.6493 ±0.0411 | 0.6461 ±0.0439 | 4.3618 ±1.0495 | 0.1819 ±0.0349 | 0.8343 ±0.017 | 0.9013 ±0.0133 |
| ResNet101V2-233 | 82.58 ±2.21 | 81.9 ±4.78 | **83.32** ±3.71 | 84.17 ±2.55 | 81.31 ±3.7 | 0.6535 ±0.0429 | 0.6515 ±0.0439 | **5.104** ±0.9811 | 0.2163 ±0.0541 | 0.8292 ±0.0254 | 0.9118 ±0.0149 |
| InceptionV3-274 | 82.73 ±2.08 | 86.57 ±2.42 | 78.6 ±2.8 | 81.33 ±2.12 | 84.52 ±2.52 | 0.6551 ±0.0419 | 0.6533 ±0.0419 | 4.1259 ±0.639 | 0.1714 ±0.0328 | 0.8385 ±0.0195 | 0.9052 ±0.0185 |
| InceptionV4-327 | 82.76 ±1.78 | 85.7 ±3.96 | 79.58 ±2.87 | 81.92 ±1.8 | 84.02 ±3.41 | 0.6561 ±0.0358 | 0.6541 ±0.0353 | 4.2716 ±0.5734 | 0.179 ±0.0465 | 0.837 ±0.0197 | 0.9092 ±0.019 |

Table 3: (continued).

| | | | | | | | | | | | |
|---|---|---|---|---|---|---|---|---|---|---|---|
| InceptionResNetV2-500 | 82.67 ±2.06 | 83.54 ±3.88 | 81.74 ±3.16 | 83.17 ±2.16 | 82.37 ±3.32 | 0.6541 ±0.0406 | 0.653 ±0.041 | 4.6886 ±0.7264 | 0.2009 ±0.0456 | 0.8329 ±0.0218 | 0.9011 ±0.0133 |
| Xception-118 | 82.48 ±2.45 | 83.16 ±5.1 | 81.75 ±2.76 | 83.08 ±1.94 | 82.16 ±4.4 | 0.6507 ±0.0487 | 0.6492 ±0.0484 | 4.6434 ±0.6571 | 0.2054 ±0.0609 | 0.8304 ±0.0276 | 0.9081 ±0.0148 |
| DenseNet121-379 | 83.88 ±0.92 | 85.85 ±1.76 | 81.75 ±0.95 | 83.49 ±0.69 | 84.35 ±1.57 | 0.6773 ±0.0186 | 0.6768 ±0.0184 | 4.7169 ±0.254 | 0.173 ±0.0211 | 0.8465 ±0.01 | 0.9158 ±0.0097 |
| DenseNet169-395 | 83.66 ±1.25 | **88.6** ±3.59 | 78.35 ±2.75 | 81.54 ±1.53 | **86.68** ±3.33 | 0.6758 ±0.0265 | 0.6717 ±0.0249 | 4.1454 ±0.4187 | **0.1446** ±0.0414 | 0.8486 ±0.0138 | 0.9123 ±0.0129 |
| DenseNet201-561 | 83.12 ±1.11 | 85.61 ±1.81 | 80.45 ±3.92 | 82.61 ±2.7 | 83.93 ±1.13 | 0.663 ±0.0221 | 0.6615 ±0.0228 | 4.5729 ±0.9885 | 0.1783 ±0.0153 | 0.8403 ±0.0073 | 0.9125 ±0.0083 |
| MobileNetV2-62 | 81.68 ±1.99 | 81.94 ±3.49 | 81.39 ±1.26 | 82.55 ±1.21 | 80.85 ±3.09 | 0.6337 ±0.0394 | 0.6332 ±0.0395 | 4.4256 ±0.3596 | 0.222 ±0.0441 | 0.8222 ±0.0218 | 0.8933 ±0.0135 |
| MobileNetV3Small-182 | 81.53 ±1.98 | 84.93 ±3.29 | 77.87 ±3.89 | 80.6 ±2.55 | 82.91 ±2.85 | 0.6315 ±0.0398 | 0.6294 ±0.04 | 3.9496 ±0.6356 | 0.1933 ±0.0386 | 0.8265 ±0.0186 | 0.896 ±0.013 |
| MobileNetV3Large-193 | 82.74 ±2.17 | 83.69 ±0.43 | 81.71 ±4.6 | 83.26 ±3.39 | 82.3 ±0.89 | 0.6548 ±0.0437 | 0.6542 ±0.0442 | 4.8573 ±1.1585 | 0.2002 ±0.0117 | 0.8344 ±0.017 | 0.9034 ±0.0094 |
| NASNetMobile-617 | 81.3 ±1.45 | 83.2 ±1.66 | 79.25 ±3.98 | 81.29 ±2.65 | 81.48 ±1.07 | 0.6261 ±0.0287 | 0.6251 ±0.0297 | 4.1452 ±0.7283 | 0.2117 ±0.0156 | 0.8219 ±0.0108 | 0.8897 ±0.0152 |
| EfficientNetB0-187 | 83.13 ±1.2 | 85.21 ±3.91 | 80.89 ±2.95 | 82.83 ±1.75 | 83.79 ±3.19 | 0.6636 ±0.0244 | 0.6618 ±0.0237 | 4.5522 ±0.6116 | 0.1817 ±0.0427 | 0.8392 ±0.0147 | 0.9094 ±0.0129 |
| EfficientNetB1-308 | 82.42 ±1.04 | 85.85 ±2.14 | 78.71 ±3.75 | 81.37 ±2.34 | 83.9 ±1.59 | 0.6492 ±0.0202 | 0.647 ±0.0214 | 4.1494 ±0.6707 | 0.179 ±0.0209 | 0.835 ±0.0074 | 0.9088 ±0.0134 |
| EfficientNetB2-316 | 82.75 ±1.4 | 84.95 ±3.41 | 80.39 ±3.02 | 82.39 ±1.91 | 83.4 ±2.69 | 0.6556 ±0.0276 | 0.6542 ±0.0279 | 4.4211 ±0.6202 | 0.1865 ±0.0379 | 0.8359 ±0.0158 | 0.9075 ±0.0082 |
| EfficientNetB3-194 | 83.46 ±0.87 | 85.15 ±4.28 | 81.64 ±2.9 | 83.4 ±1.6 | 83.9 ±3.14 | 0.6704 ±0.0157 | 0.6685 ±0.0167 | 4.7361 ±0.6283 | 0.1803 ±0.0443 | 0.8416 ±0.0144 | 0.9163 ±0.0074 |
| EfficientNetB4-384 | 82.98 ±1.31 | 87.55 ±2.2 | 78.06 ±3.76 | 81.2 ±2.33 | 85.46 ±1.76 | 0.6613 ±0.0249 | 0.6581 ±0.0268 | 4.0946 ±0.6159 | 0.1589 ±0.0233 | 0.842 ±0.0107 | 0.9138 ±0.0074 |
| EfficientNetB5-444 | 83.7 ±1.21 | 86.85 ±2.89 | 80.32 ±3.73 | 82.71 ±2.39 | 85.17 ±2.38 | 0.6752 ±0.024 | 0.6729 ±0.0245 | 4.5562 ±0.7645 | 0.1629 ±0.0303 | 0.8466 ±0.0108 | 0.9138 ±0.0161 |

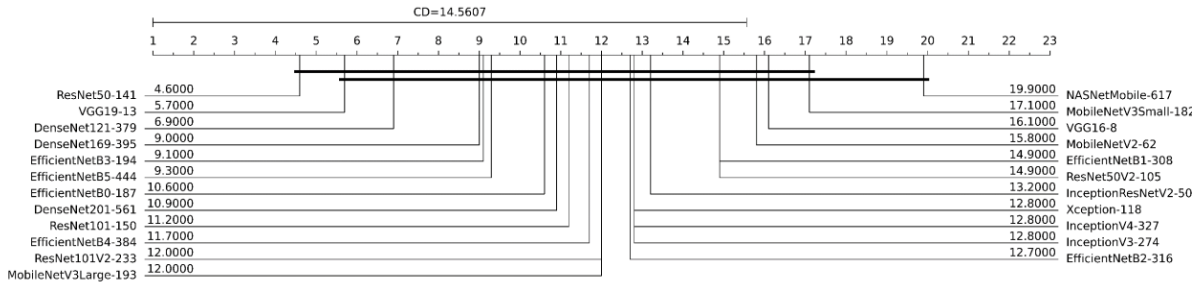

Figure 16: Accuracy critical difference diagram for the best performing configurations of the trained DCNN models. The models are ordered by best to worst average ranking from left to right. The number beside a model's name represents the average rank of the model. CD is the critical difference for Nemenyi post-hoc test. Thick horizontal line connects the models that are not statistically significantly different.

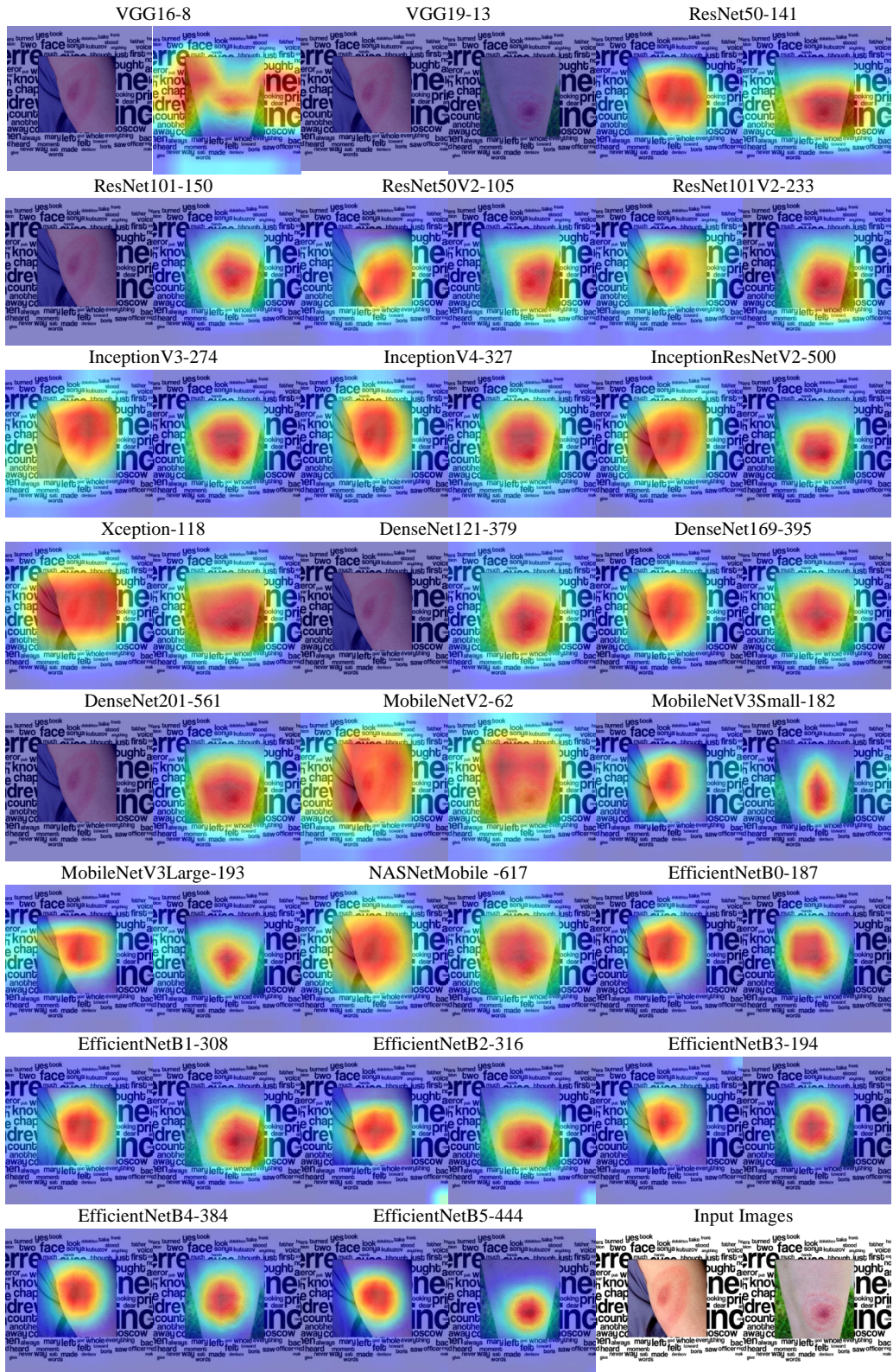

Figure 17: Grad-CAM visualization of the trained models. Input images are shown at the lower right corner.

Table 4: Complexity metrics of trained CNN models. Bold indicates the best result for each of the metrics.

| Model | Parameters (million) | FLOPs (giga) | Average training time (sec per epoch) | Disk usage (megabyte) | GPU usage (megabyte) | Average inference time (sec per image) | Input shape |
|---|---|---|---|---|---|---|---|
| VGG16-8 | 14.72 | 30.7 | 111 | 146.24 | 565 | 0.0426 | 224x224x3 |
| VGG19-13 | 20.02 | 39 | 164 | 216.03 | 565 | 0.0431 | 224x224x3 |
| ResNet50-141 | 23.59 | 7.75 | 113 | 268.64 | 821 | 0.0484 | 224x224x3 |
| ResNet101-150 | 42.66 | 15.2 | 123.33 | 378.41 | 821 | 0.0539 | 224x224x3 |
| ResNet50V2-105 | 23.57 | 6.99 | **76** | 258.84 | 821 | 0.0464 | 224x224x3 |
| ResNet101V2-233 | 42.63 | 14.4 | 152 | 429.45 | 821 | 0.0599 | 224x224x3 |
| InceptionV3-274 | 21.8 | 11.5 | 133 | 246.94 | 821 | 0.0540 | 224x224x3 |
| InceptionV4-327 | 41.18 | 24.6 | 223.33 | 424.93 | 1333 | 0.0735 | 299x299x3 |
| InceptionResNetV2-500 | 54.34 | 26.4 | 281.33 | 588.12 | 1333 | 0.0958 | 299x299x3 |
| Xception-118 | 20.86 | 16.8 | 243.33 | 238.48 | 821 | **0.0392** | 299x299x3 |
| DenseNet121-379 | 7.04 | 5.7 | 140.67 | 78.7 | 437 | 0.0673 | 224x224x3 |
| DenseNet169-395 | 12.64 | 6.76 | 130 | 128.59 | 565 | 0.0686 | 224x224x3 |
| DenseNet201-561 | 18.32 | 8.63 | 182.67 | 198.7 | 565 | 0.0840 | 224x224x3 |
| MobileNetV2-62 | 2.26 | 0.613 | 78 | 24.02 | **341** | 0.0429 | 224x224x3 |
| MobileNetV3Small-182 | **1.53** | **0.174** | 81 | **17.8** | **341** | 0.0444 | 224x224x3 |
| MobileNetV3Large-193 | 4.23 | 0.564 | 86.33 | 48.34 | 373 | 0.0444 | 224x224x3 |
| NASNetMobile -617 | 4.27 | 1.15 | 152 | 50.74 | 373 | 0.0741 | 224x224x3 |
| EfficientNetB0-187 | 4.05 | 0.794 | 87 | 46.59 | 373 | 0.0523 | 224x224x3 |
| EfficientNetB1-308 | 6.58 | 1.41 | 158.33 | 75.89 | 437 | 0.0546 | 240x240x3 |
| EfficientNetB2-316 | 7.77 | 2.04 | 210 | 89.53 | 437 | 0.0565 | 260x260x3 |
| EfficientNetB3-194 | 10.79 | 3.74 | 143 | 117.6 | 565 | 0.0648 | 300x300x3 |
| EfficientNetB4-384 | 17.68 | 8.97 | 431 | 202.2 | 565 | 0.0614 | 380x380x3 |
| EfficientNetB5-444 | 28.52 | 20.9 | 771 | 325.21 | 821 | 0.0659 | 456x456x3 |

## 4. Discussion

The experimental result described above makes it evident that CNNs have great potential to be used for Lyme disease pre-scanner application. Figure 18 shows a bubble chart reporting model accuracy vs FLOPs. The size of each bubble represents the number of parameters of the model. This figure serves as a guideline for selecting models based on complexity and accuracy. It can be seen from the figure that EfficientNetB0-187 is a good choice with reasonable accuracy for resource-constrained mobile platforms. EfficientNetB0-187 also showed good results in Grad-CAM visualization. If resource constraint is not a problem, then RestNet50-141 can be used for the best accuracy.

Even the lightweight EfficientNetB0-187 model showed good performance, and it can be directly deployed in mobile devices without requiring an internet connection for processing the lesion image in a remote server. It can help people living in remote areas without good internet facilities with an initial assessment of the probability of Lyme disease.

For this study, we utilized images from the internet alongside images collected from several hospitals in France. This approach was inspired by related studies on skin lesion analysis. Although a portion of images in our dataset was collected from the internet the annotation of the dataset is reliable because we ignored the online labels, and all the images were reannotated by expert dermatologists and infectiologists.

Existing works including this study on AI-based Lyme disease analysis only utilize images but including patients' metadata can be a great way of strengthening the analysis. We trained the CNNs with whole images without EM lesion segmentation. The effect of the EM lesion segmentation on the predictive performance of CNNs can be an interesting study. Another limitation is that dark-skinned samples are underrepresented in our dataset. A limited number of samples in the dataset with skin hair artifact over the EM lesion is also a concern. Although hair removal algorithms can be used for removing skin hair it will increase the computational complexity of the real-time mobile application. A better alternate can be augmenting the training dataset with artificial skin hair.

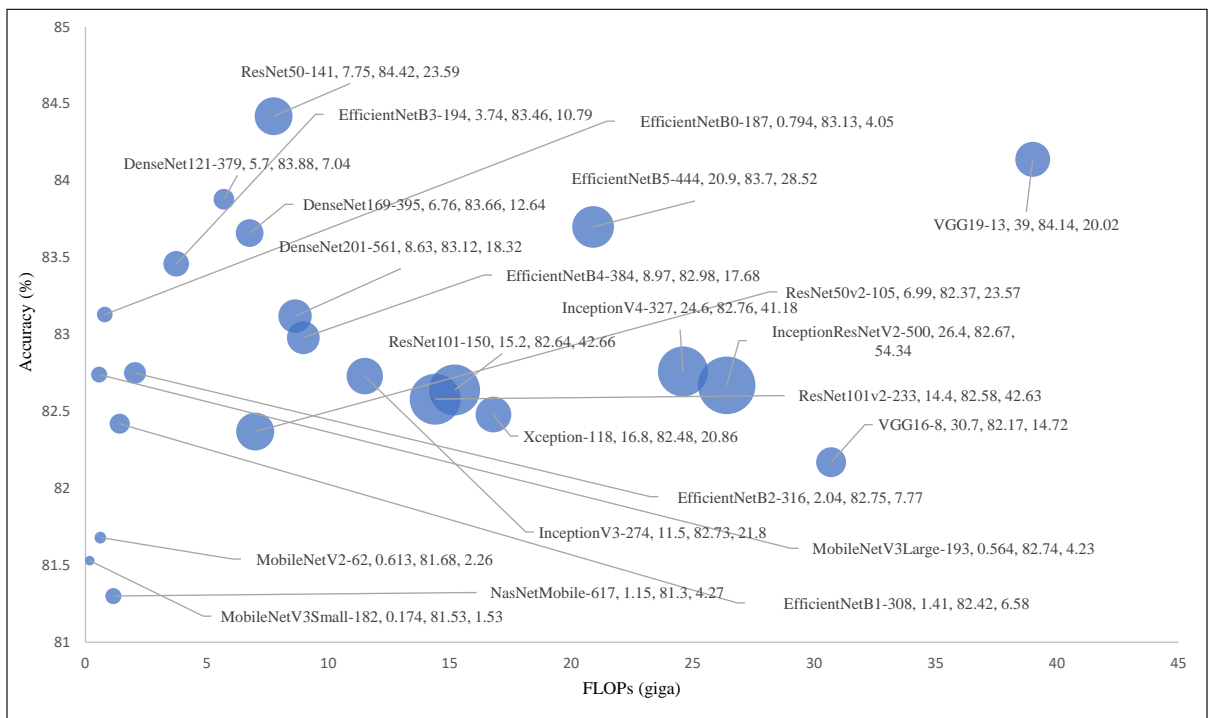

Figure 18: Bubble chart reporting model accuracy vs floating-point operations (FLOPs). The size of each bubble represents number of model parameters measured in millions unit. Beside each model name the three values represent FLOPs, accuracy, and model parameters, respectively.

## 5. Conclusion

In this study, we benchmarked and extensively analyzed twenty-three well-known CNNs based on predictive performance, complexity, significance tests, and explainability using a novel Lyme disease dataset to find out the effectiveness of CNNs for Lyme disease diagnosis from EM images. To improve the performance of CNNs we utilized HAM10000 dataset with ImageNet pre-trained models for transfer learning. We also provided guidelines for model selection. We found that even the lightweight models like EffiicentNetB0 performed well suggesting the application of CNNs for Lyme disease pre-scanner mobile applications which can help people with an initial assessment of the probability of Lyme disease and referring them to expert dermatologist for further diagnosis. Resource intensive models like ResNet50 can be effective for building computer applications to assist non-expert practitioners with identifying EM. We also made all the trained models publicly available, which can be utilized by others for transfer learning and building pre-scanners for Lyme disease.

**Declaration of Competing Interest**

The authors have declared no conflict of interest.

**Acknowledgments**

This research was funded by the European Regional Development Fund, project DAPPEM –AV0021029. The DAPPEM project («Développement d'une APPlication d'identification des Erythèmes Migrants à partir de photographies»), was coordinated by Olivier Lesens and was carried out under the Call for Proposal 'Pack Ambition Research' from the Auvergne-Rhône-Alpes region, France. This work was also partially funded by Mutualité Sociale Agricole (MSA), France.

**Data Statement**

We can share images collected from the internet and the corresponding labels assigned by the doctors subject to an agreement that the images will not be made public because we do not have permission from the owners of the images. Interested researchers can send a request to dappem-project@inrae.fr for the data. The images collected from the hospital are not shareable because of the confidentiality agreement signed with the patients.

**Remarks:**

This is a preprint. For the published article please refer to:
https://doi.org/10.1016/j.cmpb.2022.106624

Cite this research article as:
S.I. Hossain, J. de Goër de Herve, M.S. Hassan, D. Martineau, E. Petrosyan, V. Corbin, J. Beytout, I. Lebert, J. Durand, I. Carravieri, A. Brun-Jacob, P. Frey-Klett, E. Baux, C. Cazorla, C. Eldin, Y. Hansmann, S. Patrat-Delon, T. Prazuck, A. Raffetin, P. Tattevin, G. Vourc'h, O. Lesens, E.M. Nguifo, Exploring convolutional neural networks with transfer learning for diagnosing Lyme disease from skin lesion images, Comput. Methods Programs Biomed. 215 (2022) 106624. https://doi.org/10.1016/j.cmpb.2022.106624.